# Enhanced Electromechanical Response in Defective Sm and Nd Co-doped Ceria

Ahsanul Kabir[1], Jacob R. Bowen[1], Maxim Varenik[2], Igor Lubomirsky[2], Vincenzo Esposito[1*]

[1]Department of Energy Conversion and Storage, Technical University of Denmark (DTU), Kgs. Lyngby 2800, Denmark

[2]Department of Materials and Interfaces, Weizmann Institute of Science (WIS), Rehovot 76100, Israel

*Corresponding Authors: E-mail: ahsk@dtu.dk, vies@dtu.dk

# Abstract

Highly oxygen defective cerium oxide, *e.g.* Gd-doped ceria is a sustainable non-classical electrostrictor with electromechanical properties that are superior to lead-based piezoelectric metal oxides. Here, we report electrostriction in co-doped ceria (Sm, Nd) with a nominally low short-range vacancy-dopant association energy. Such a strategy results in a higher electrostrictive strain coefficient ($M_{33}$), up to $10^{-17}\,\text{m}^2/\text{V}^2$ at lower-frequencies, and unexpected electromechanical strain saturation and relaxation effects. These outcomes support the hypothesis that electrostriction is strongly influenced by the local environment of oxygen vacancy and by the ionic migration blocking factors built-in the microstructure.

## 1. Introduction

Electrostrictive materials are subjected to a wide variety of technological applications as actuators in mechanics, electronics and biomedical devices [1][2][3]. Electrostriction is the electromechanical response that arises in electrostrictors as the second-order elastic deformation under an applied electric field [1][4]. It is present in all dielectrics regardless of crystal structure and is described by a fourth-ranked tensor that follows the relationship:[5]

$$x_{ij} = M_{ijmn} E_m E_n \tag{1}$$

where $M_{ijmn}$ is the electrostrictive field related strain coefficient and $E$ is the external electric field. The values of $M_{ijmn}$ typically range (for inorganic materials) from about $10^{-21} - 10^{-16}$ $m^2/V^2$ depending on both the dielectric permittivity and elastic modulus of the materials [1]. However, the electrostrictive strain is not always quadratic, as the dielectric constant often increases non-linearly with the electric field, especially in high permittivity materials. Accordingly, Newnham *et al.* proposed to characterize electrostrictive properties by means of polarization electrostriction coefficient $Q_{ijmn}$, described as: [1]

$$x_{ij} = Q_{ijmn} P_m P_n \tag{2}$$

$$P_m = E_n \eta_{mn} \tag{3}$$

Here, $P$ is the dielectric polarization and $\eta_{mn}$ is the dielectric susceptibility tensor. Furthermore, to compare electrostrictive materials with different crystallographic symmetry a universal hydrostatic electrostriction polarization coefficient ($Q_h$) has been considered, which is associated with the dielectric permittivity ($\varepsilon$) and inverse Young's modulus ($S$) in an empirical relation [6]:

$$|Q_h| \approx 2.37 \cdot (S/\varepsilon\varepsilon_0)^{0.59} \tag{4}$$

Recently, it has been shown that thin films of defective cerium oxide *e.g.* 20 mol% Gd-doped ceria (GDC), ($\varepsilon^{GDC} \approx 30$ and $Y^{GDC} \approx 200$ GPa) exhibit giant electrostrictive strain coefficient ($M_{33}$) around $\sim 6.5 \times 10^{-18}$ $m^2/V^2$ at 0.1 Hz [7]. The estimated $|Q_h|$ reveals at least two orders of magnitude larger than the Newnham scaling law in Eqn 4. Similar results have been reported for highly defective GDC and bismuth oxide-based bulk ceramics [6][8][9]. The atomistic mechanism functioning this non-classical type of electrostriction is due to the presence of electroactive Ce-O bonds in the crystal lattice, as explained by Lubomirsky and co-workers [7][10][11]. Based on extended X-ray absorption fine structure (EXAFS) and X-ray absorption near edge structure (XANES) experiments, it has been observed that oxygen vacancies ($V_O^{\cdot\cdot}$) in the lattice create a small population of distorted $Ce_{Ce}$-$7O_o$-$V_O^{\cdot\cdot}$ units, which consists of elongated Ce-$V_O^{\cdot\cdot}$ and contracted Ce-O bonds in comparison with the Ce-O bond in the Ce-$8O_o$ unit. This results in asymmetric local charge configuration and anisotropic local dipolar elastic strain around the lattice [10][12]. Hence, upon application of an electric field, these electroactive bonds change dynamically and become ordered, inducing a large macroscopic electromechanical displacement [10][12]. In other investigations, it has been explained that $M_{33}$ in GDC bulk ceramics exhibits strain saturation and relaxation in response to the applied electric field and frequency, respectively [8][9]. To date, the electrostriction effect in defective ceria is investigated only with pure and Gd-doped ceria. Therefore, it is important to understand the role of other types of dopants *i.e.* the physical correlation between dopant associated oxygen vacancy concentration and the electrostriction.

In the present work, we aim to optimize the electrostrictive properties by tuning the configuration of oxygen vacancy in the lattice by substituting Gd (**see Fig. 1.a**) with a co-dopant combination of Sm/Nd (1:1 ratio). Based on the computational prediction reported in previous work [13], a equimolar co-doped system with an effective atomic number around 61 exerts lesser interaction between dopant and oxygen vacancy compared to a singly doped ceria compound *i.e.* oxygen vacancies show no site preferences, eventually resulting in an enhanced configurational entropy [13][14][15]. Although difficult to formalize, the configuration of oxygen vacancy can be macroscopically characterized by electrochemical impedance spectroscopy [16][17]. This technique can resolve the interlinked effects between the dopants, oxygen vacancies concentration, and microstructure by separating the overall macroscopic electrochemical response into defined domains of electrical impedance, *i.e.* of electrochemical capacitance and ionic migration blocking effects [18][19][20]. The resolution of the ionic blocking factors is here correlated with the electromechanical response, in an attempt to clarify the interlinked effects between oxygen vacancy concentration and their configuration in doped ceria.

## 2. Experimental Procedure

Nanometer-scale Sm and Nd-co-doped ceria (CDC) powders with a composition of $Ce_{1-x}Sm_{x/2}Nd_{x/2}O_{2-x/2}$ (where $x$ = 0.01, 0.05, 0.1, and 0.15 referred to as CDC-1, CDC-5, CDC-10, and CDC-15, respectively) were synthesized by the co-precipitation method [18]. Metal nitrate salts (Sigma-Aldrich, USA) of the respective element were dissolved in a stoichiometric ratio in deionized water (0.1 M solution). Afterward, N-methyl-diethanolamine was slowly added to the solution, keeping the pH ≈10. The precipitates were formed immediately and were kept overnight under slow stirring. Precipitates were then centrifuged and washed several times with ethanol, followed by calcination at 500 °C for 2 h. Hereafter, the dried hard agglomerated powders were ball-milled in ethanol at 50 rpm for 10 h and subsequently dried at 120 °C. Finally, the powders were softly ground in an agate mortar pestle and sieved through a 150 µm mesh. The powders were then cold-pressed as pellets (12 mm diameter, 1-1.2 mm thickness) with a uniaxial pressure of 150 MPa and sintered at 1450 °C for 10 h. The experimental density of the samples was measured by the Archimedes method in distilled water. The crystallographic phase composition was verified by X-ray powder diffraction (XRD) technique (Bruker D8, Germany). The microstructure was characterized by a high-resolution scanning electron microscopy (SEM, Zeiss Merlin, Germany). The electrical conductivity of the samples is examined in the range of ~250-500 °C in 25 °C intervals in the air by impedance spectroscopy (Solarton 1260, UK) in a frequency range of 0.01 Hz to 10 MHz with an applied 100 mV alternating current signal. For the electromechanical response, the strain was evaluated with a proximity sensor of the capacitive type (Lion Precision, USA) with lock-in detection [8][9].

## 3. Results and Discussion

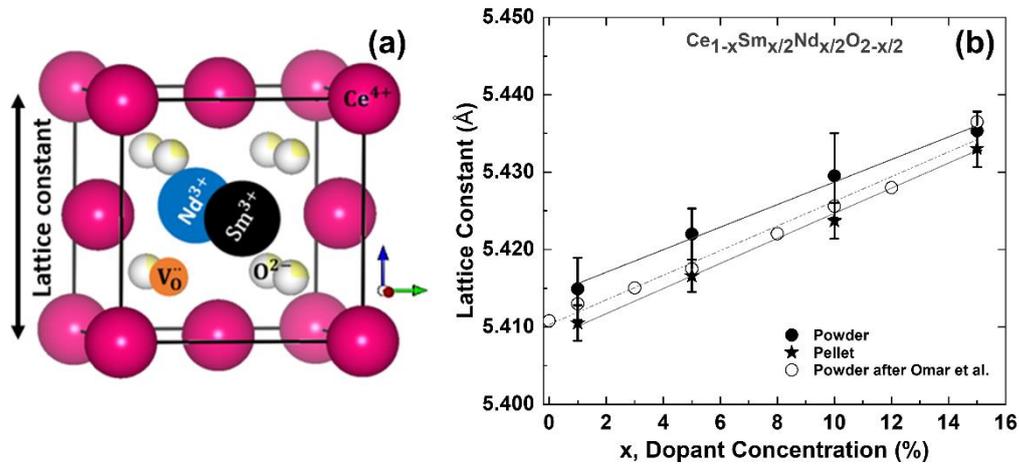

**Figure 1:** (a) Schematic representation of a typical unit cell of ceria lattice in the presence of trivalent co-dopants. (b) Lattice constant as a function of total dopant concentration (mol%) where x = 0.01−0.15. Results are compared with the data of Omar *et al.* [14].

The X-ray diffraction (XRD) patterns are shown in Fig. **S1** and **S2**, reveal a single-phase cubic fluorite structure. Considering the detection limits of the techniques (5-7 % in volume), no other impurity diffraction peaks were evidenced. Moreover, the increment of dopant concentration shifted the featured peaks *e.g.* (111) toward lower 2θ angles, confirming the expansion of the lattice parameter with respect to dopant concentration. The estimated lattice constant of both the starting powders and sintered pellets as a function of total dopant content is illustrated in **Fig. 1.b** which shows that the lattice constant increases linearly with the concentration of dopant, according to Vegard's rule [21]. A similar linear trend was demonstrated by Omar *et al.* in Sm, Nd-co-doped ceria system [14][15]. Moreover, other reports describe that the lattice constant of the singly doped-ceria system follows a quadratic expression with dopant concentration where the coefficient of the second-order is rendered as oxygen vacancy ordering [22][23][24]. The absence of second-order term underlines that dopant associated short-range oxygen vacancy ordering in the CDC sample is relatively less in comparison with singly-doped ceria, as demonstrated in other previous work [14]. In addition, lattice constant decreases considerably for the sintered pellet. This is due to the fact that long-term thermal treatment at higher temperatures i.e. 1450 °C for 10 hours initiates the lattice defect annihilation during the sintering and densification process by the mass transport mechanism [25][26][27].

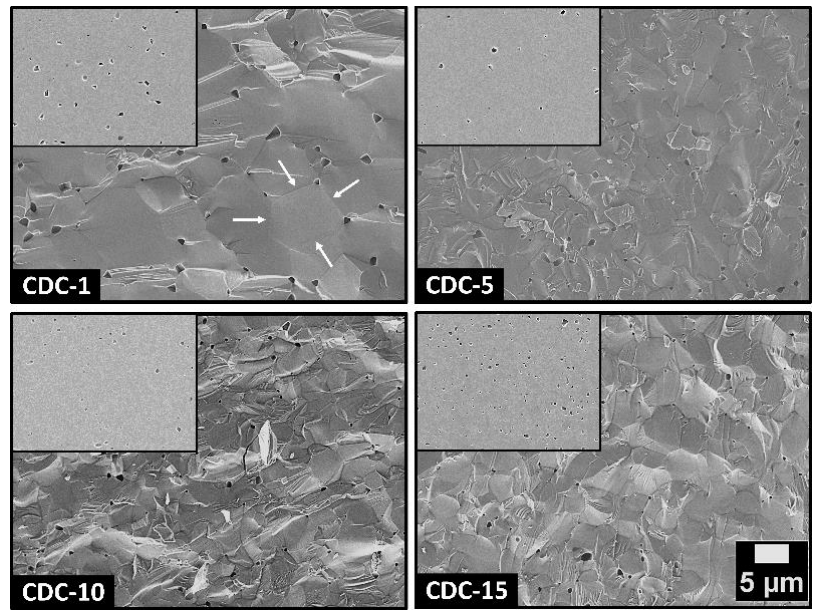

**Figure 2:** The SEM micrographs of the cold-fractured and polished cross-sectional surface of the sintered pellets. All samples have the same scale bar = 5 µm.

The relative density of all sintered pellets was found to be above 95%, in agreement with the observed microstructural results in **Fig. 2** where only a few percents of isolated residual porosity was observed. The average grain size of all samples as measured by the linear intercept method [28] ranges in-between 3-6 µm, showing a reducing trend with increasing dopant content. These grains are relaxed, having a small residual grain boundary curvature (white arrows).

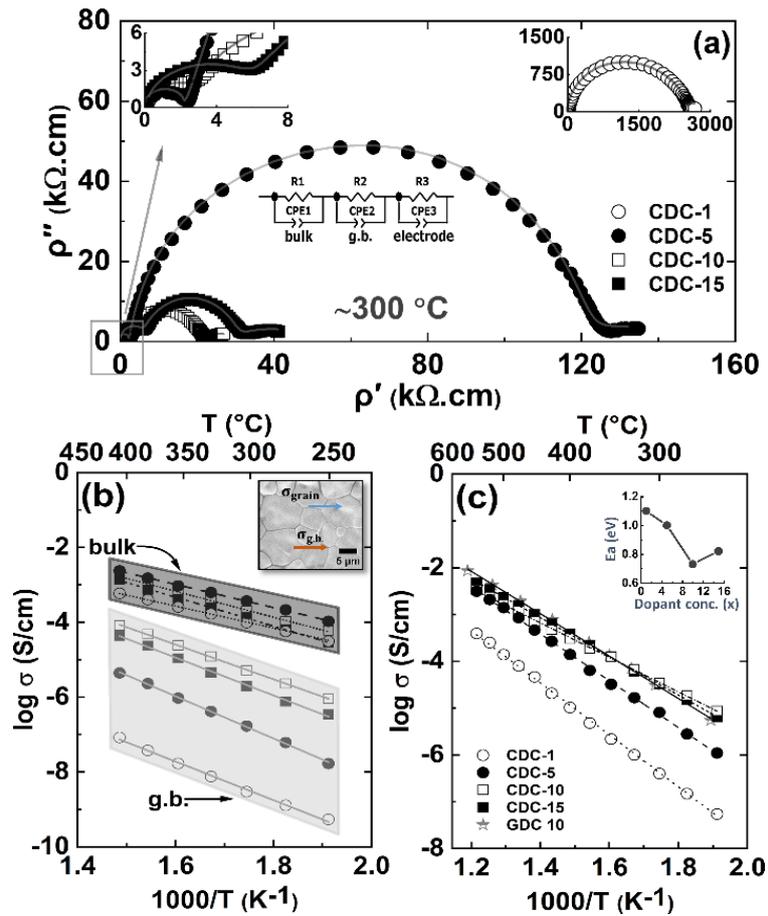

**Figure 3:** (a) Geometry normalized Nyquist plots (ρ' vs ρ") examined in the air at 300 °C. Arrhenius plot for the estimation of (b) bulk and grain boundary conductivities, (c) total electrical conductivities, of sintered CDC pellets. The GDC-10 sample (sintered at 1450 °C for 10h) is taken from ref. [9].

The characterization of the ionic electrical properties at low temperatures identifies transport mechanisms of oxygen defect in the materials as a result of association and blocking effects at the bulk/grain boundary. Particularly, electrochemical impedance spectroscopy at moderately high frequency allows separating such contributions. The results are presented in **Fig. 3**, as analyzed by the method described in ref. [16][17]. It is worth remarking that, despite being generally associated with the grain boundaries, the ionic blocking effect can also arise from other chemical and microstructural factors *e.g.* porosity, nanodomain formation, impurity/dopant segregation, etc. [29][30][31]. In **Fig. 3a**, the geometry normalized Nyquist plots at 300 °C exhibit two well-defined semicircles that correspond to high and intermediate frequency attributed bulk and grain boundary resistance of the co-doped samples, respectively. The low-frequency arc refers to the electrolyte-electrode polarization and it is not relevant in this work. These Nyquist plots were fitted by parallel equivalent circuits of RQ elements where R and Q are denoted respectively as resistor and constant phase element. As observed, the bulk resistance differs moderately among the samples, however, a prominent deviation is measured in grain boundary contribution, especially for CDC-1 and CDC-5 compound. Both of them display a grain boundary blocking factor ($\alpha_{gb}$) about ≈0.98 at 300 °C, which is the ratio of grain boundary resistance to the total resistance ($\alpha_{gb} = \frac{R_{gb}}{R_{bulk}+R_{gb}}$) [32]. The high blocking effect also decreases the characteristic frequency response of the grain boundary (see **Table 1**). The Arrhenius plot in **Fig. 3.b** illustrates the bulk (grain) and (specific) grain boundary electrical conductivity ($\sigma = 1/\rho'$) for all samples over the temperature ranges of ~250-400 °C. Noteworthy is the pronounced difference between these two conduction mechanisms, with the value of bulk conductivity typically exceeds a factor of two-four larger than grain boundary. When comparing bulk conductivity between the samples, conductivity increases with dopant concentration for low doping (<5 mol%) and falls off for the highly doped system. Such an outcome can be attributed to the fact that the highly doped compositions, *e.g.* CDC-10 and higher, possibly develop a cation-defect association interaction, which not only minimizes free oxygen vacancy concentration but also decreases the number of low energy migration paths for diffusion by forming high migration energy barriers [13][33]. This finding is in agreement with the enhanced activation energy value for bulk (see **Table 1**). A preceding experimental study by Omar and co-workers for the grain ionic conductivity of $Ce_{1-x}Sm_{x/2}Nd_{x/2}O_{2-x/2}$ exhibits a parallel trend [14]. The hypothesis of dopant-defect association in oxygen defective cerium oxides is well-consolidated for Gd, Nd, Sm- and other substitutional doping [34]. The ion blocking effect was assumed to be dominated by the grain boundary, which is strongly influenced by nominal dopant concentration. The grain boundary conductivity indicates an incremental trend until 10 mol% doping then slightly decreases for the CDC-15 sample. According to most literature reports, grain boundary conductivity increases with dopant concentration because of the decreasing trend of space charge potential as a function of nominal dopant content [35][36]**.** However, the in-depth mechanism of blocking effect is difficult to explain exclusively via electrochemical measurements and it can also be attributed to dopant segregation effects at the grain boundary, as mentioned above [26][27]. From the Arrhenius plot in **Fig. 3b**, the activation energy for grain boundary conductivity is estimated at around 1.0 eV, which is typical for the ceria compound. The total conductivity is affected by the co-operative effects of both bulk and grain boundary and the plot in **Fig. 3c** displays that total electrical conductivity increases with dopant concentration. The inset of this plot represents the activation energy for total conductivity, exhibiting a higher value for the sample with low dopant concentration. Such an outcome is possibly associated with the low dopant-vacancy ordering and to the large ion blocking barrier effects measured at the grain boundary. The total conductivity of CDC-10, CDC-15, and reference GDC-10 sample have comparable values, an order of magnitude higher than CDC-5, particularly in the low-temperature regime (<375 °C). In summary, the impedance results show that at a fixed thermal treatment *i.e.* 1450 °C for 10h in this experiment results in a large ion blocking effect for dilute doping. The blocking effect have a tendency to scale-down with increasing nominal dopant content (see **Table 1**). A direct comparison with GDC-10 (sintered at same condition), the CDC-10 sample represents a reduced blocking effect.

**Table 1:** The activation energy, relaxation frequency, and grain boundary blocking effect data of CDC ceramics, in comparison with the GDC-10 sample sintered at 1450 °C for 10h [9].

| Sample ID | $E_{grain}$ (eV) | $E_{g.b.}$ (eV) | ~ $f_{grain}$ (Hz) | ~ $f_{g.b.}$ (Hz) | $\alpha_{g.b.}$ at 300 °C |
|---|---|---|---|---|---|
| CDC-1 | 0.58 | 1.0 | $1 \times 10^6$ | $1.5 \times 10^1$ | 0.99 |
| CDC-5 | 0.60 | 1.10 | $2 \times 10^6$ | $3 \times 10^2$ | 0.98 |
| CDC-10 | 0.68 | 0.90 | $1 \times 10^5$ | $1 \times 10^3$ | 0.80 |
| CDC-15 | 0.76 | 1.0 | $8 \times 10^4$ | $1 \times 10^3$ | 0.75 |
| GDC-10 | 0.65 | 0.95 | $2 \times 10^5$ | $1.5 \times 10^3$ | 0.90 |

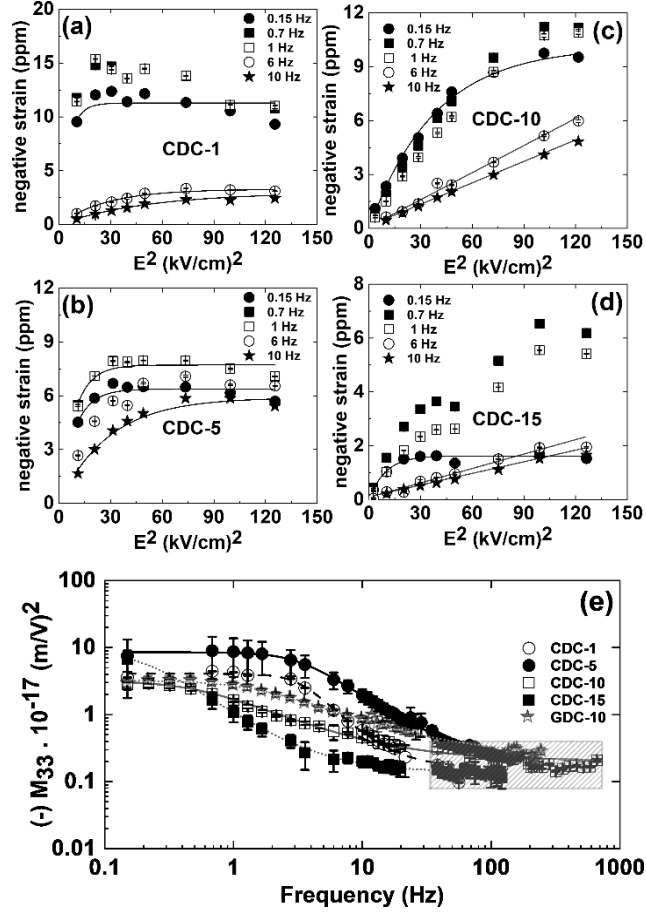

**Figure 4:** (a)-(d) Electrostrictive negative strain as a function of applied electric field square at frequencies between 0.15-10 Hz, displaying saturation of strain at a higher electric field. The error bar is smaller than the symbol size (e) The relaxation phenomena of electrostriction strain coefficient ($M_{33}$) as a function of applied field frequencies (below 50 Hz). The results are compared with the GDC-10 sample, sintered at 1450 °C for 10h [9].

The electrostrictive strain with a response to the square of the applied electric field at selected frequencies (up to 10 Hz) is demonstrated in **Fig. 4.a-d**. As expected, all CDC samples exhibit a negative longitudinal strain at the second harmonic of the electric field, in agreement with the previously published reports of GDC bulk materials [8][9]. In general, the observed strain ($u_{33}$) value tends to saturate and unexpectedly in few cases decrease at higher electric fields. The strain saturation behavior is empirically fit to the following equation:

$$u_{33} = M_{33} \cdot E_{sat}^2 \cdot \left[1 - \exp(-E^2/E_{sat}^2)\right] \qquad (5)$$

Here, $M_{33}$ is the electrostriction strain coefficient and $E_{sat}$ is the saturation electric field.

Additionally, the figures also display that with increasing frequency, the magnitude of electromechanical strain declines significantly. Interestingly, both the CDC-10 and CDC-15 samples follow the linear trend of electrostriction strain vs $E^2$ above 6 Hz in the entire range of investigated electric field. The lack of strain saturation at these frequencies could be associated with the low blocking barrier effect, formed in these compositions. Despite inconclusive, such experimental results are consistent with previous [9] and other unpublished results. For example, the microcrystalline bulk GDC revealed no strain saturation when the blocking factor is noticeably reduced *i.e.* around ≈0.65 [9]. The electrostriction strain coefficient ($M_{33}$) as a function of applied field frequencies are represented in **Fig. 4.e**. All samples display non-ideal Debye type relaxation and are fitted with the following relation:

$$M_{33}(f) = \frac{M_{33}^0}{\sqrt{1+(\tau.f)^{2+\alpha}}} + M_{33}^\infty \qquad (6)$$

Where $M_{33}^0$ and $M_{33}^\infty$ are the electrostriction coefficient at low and high frequencies, respectively. $\tau$ and $\alpha$ are denoted as relaxation time and non-ideality factor, correspondingly. The fitting parameters from Eqn (6) are presented in **Table 2**. In low-frequency regime *e.g.* below 10 Hz, both the CDC-1 and CDC-5 samples exhibit significantly higher $M_{33}$ values than of highly doped CDC-10 and CDC-15. The former samples have low oxygen vacancy concentration but a high ion blocking factor compared to the latter. In addition to that, the low-frequency $M_{33}$ and blocking barrier value of the GDC-10 sample lies in between. From this perspective, the current findings emphasize that $M_{33}$ value in the low-frequency regime is strongly dominated by the blocking barrier effect developed in the materials. The ion blocking barriers are usually controlled by both microstructural features (grain size) and especially by the effective dopant distribution, as also shown in the previous study [9]. Moreover, at higher frequencies, *e.g.* above 50 Hz, the magnitude of $M_{33}$ is firmly steady (shaded box in **Fig. 4.e**), ranging in between $0.1-0.3 \times 10^{-17}$ m²/V² for all samples and is independent of both oxygen vacancy concentration and configuration. Above all, $M_{33}$ values for all the samples at the high-frequency regime are estimated in the order of ≈$10^{-18}$ m²/V² that is equivalent to $|Q_h|$ ≈6.5 $m^4/C^2$, a value that is still one order of magnitude larger than what Newnham's classical model predicts.

Table 2: Fitting parameters in Eqn (6), in comparison with GDC-10 ceramics [9].

| Sample ID | (−) $M_{33}^0$, $10^{-17}$ m²/V² | (−) $M_{33}^\infty$, $10^{-17}$ m²/V² | τ, S | α |
|---|---|---|---|---|
| CDC-1 | 6.9 ± 0.23 | 0.15 ± 0.02 | 0.45 ± 0.03 | 2.0 ± 0.3 |
| CDC-5 | 8.3 ± 1.3 | 0.20 ± 0.06 | 0.30 ± 0.01 | 0.70 ± 0.1 |
| CDC-10 | 2.9 ± 0.05 | 0.19 ± 0.01 | 1.8 ± 0.1 | −0.3 ± 0.1 |
| CDC-15 | 3.3 ± 0.2 | 0.13 ± 0.01 | 2.4 ± 0.05 | 0.8 ± 0.06 |
| GDC-10 | 3.0 ± 0.1 | 0.20 ± 0.04 | 0.7 ± 0.05 | −0.3 ± 0.1 |

## 4. Conclusion

In this work, oxygen defective ceria compounds were synthesized by a co-doping strategy with a composition of $Ce_{1-x}Sm_{x/2}Nd_{x/2}O_{2-x/2}$ where x = 0.01–0.15 and sintered at higher temperatures. The equimolar co-doping minimizes the short-range oxygen vacancy ordering in the lattice. Moreover, all samples develop a different ion blocking effect which in turn depends on the nominal oxygen vacancy concentration and their configuration. These materials exhibit non-classical electrostriction with an unusual strain saturation and relaxation mechanisms as a function of applied electric field amplitude and frequency. For instance, a large electrostrictive coefficient is reported at low frequency (ca. $10^{-17}\,m^2/V^2$) under an applied electric field below 5 kV/cm. In particular, the electrostriction value at the low-frequency regime maintains a strict relationship with oxygen vacancy configuration at the blocking barrier. Electrostriction at high-frequency regime expresses neither concentration nor configuration dependency but a steady value. In view of this, we conclude that low ion blocking barrier effects are indispensable to have a constant $M_{33}$ value for a wide range of frequencies.

## Acknowledgments

This research was supported by DFF-Research project grants from the Danish Council for Independent Research, Technology and Production Sciences, June 2016, grant number 48293 (GIANT-E) and the European H2020-FETOPEN-2016-2017 project BioWings, grant number 801267.